\documentclass[prl,twocolumn,a4paper]{revtex4}
\usepackage{graphicx}% Include figure files
\usepackage{float}
\usepackage[latin1]{inputenc}
\usepackage{natbib}

\begin{document}
\title{The Nuclear Spin Environment in Lateral GaAs Spin Valves}
%The maximum title length is 15 words.

\author{Dominikus K\"{o}lbl}
\author{Dominik M. Zumb\"{u}hl}
\email{dominik.zumbuhl@unibas.ch} \affiliation{Department of Physics, University of Basel, Klingelbergstrasse 82,
CH-4056 Basel, Switzerland}
\author{Andreas Fuhrer}
\author{Gian Salis}
\author{Santos F. Alvarado}
\affiliation{IBM Research, Z\"{u}rich Research Laboratory, S\"{a}umerstrasse 4, 8803 R\"{u}schlikon, Switzerland}
\date{\today}% It is always \today, today,

\begin{abstract}
%The abstract is typically 150 words and is unreferenced; it contains a brief account of the background and rationale of the
%work, followed by a statement of the %main conclusions introduced by the phrase "Here we show" or some equivalent.
\textbf{The spin degree of freedom in solids offers opportunities beyond charge-based electronics and is actively
investigated for both spintronics and quantum computation. However, the interplay of these spins with their native
environment can give rise to detrimental effects such as spin relaxation and decoherence. Here, we use an
all-electrical, lateral GaAs spin valve to manipulate and investigate the inherent nuclear spin system. Hanle
satellites are used to determine the nuclear spin relaxation rates for the previously unexplored temperature range down
to 100 mK, giving T$_1$ times as long as 3 hours. Despite metallic temperature dependence of resistivity, the observed
relaxation rates show a sub-linear temperature dependence. This contrasts the Korringa relaxation mechanism observed in
metals but is not inconsistent with hyperfine-mediated relaxation in a disordered, interacting conductor not far from
the metal-insulator transition. We discuss possible relaxation mechanisms and further investigate inhomogeneities in
the nuclear spin polarization.}

\end{abstract}
\maketitle

%An introduction (without heading) of up to 500 words of referenced text expands on the background of the work (some overlap
%with the summary is acceptable), followed by a concise, focused account of the findings, ending with one or two short paragraphs of discussion.

Spintronics is a promising vision for next generation electronics \cite{Wolf2001}, initiated by the spin transistor
proposed by Datta and Das \cite{Datta1990,Loss2003}. Spin-based phenomena have become building blocks for both logic
and memory devices \cite{Kikkawa1997,Behin2010}. Furthermore, the proposal of an electron-spin quantum computer
\cite{Loss1998} has opened up a very active field of research aimed at understanding and controlling spins in
semiconductor quantum dots \cite{Hanson2007}, where the nuclear spin bath of the host material has been identified to
be of major importance. The interplay of the electron and nuclear spin systems in GaAs nanostructures leads to many
interesting effects, including dynamic nuclear polarization (DNP) \cite{Wald1994, Petta2008}, intricate mutual feedback
\cite{Ohno2004, Vink2009, Latta2009}, electron spin decoherence \cite{Bluhm2010}, and further allows coherent
manipulation of the nuclear system \cite{Yusa2005}. Weak coupling leads to long nuclear relaxation and coherence times
which makes the nuclear spins themselves interesting for logic and memory applications.

Here, we use a lateral spin valve device \cite{Johnson1985,Jedema2002} as a model system to investigate the GaAs
nuclear spin environment. A spin valve offers an efficient way to inject, manipulate and detect electron spins in GaAs
\cite{Lou2007,Salis2009,Ciorga2009}. Electron spin polarization can be transferred by a DNP process \cite{Lampel1969,
Paget1977,Kawakami2001,Strand2003} to the nuclear system, mediated by the hyperfine contact interaction via spin
flip-flop processes. The spin valve thus allows us to manipulate and investigate the native nuclear spin environment by
all-electrical means, rather than using optics \cite{Meier}, making easily accessible the low-temperature regime $T\ll
1\,K$ which was previously not explored. The nuclear spin polarization acts back on the electron spins as an effective
Overhauser field $B_{N}$ \cite{Overhauser1953} causing electron spin precession. This results in clear features in
spin transport measurements, including a depolarization peak \cite{Lou2007,Salis2009,Chan2009} and satellite
peaks in the Hanle geometry when the applied magnetic field cancels the internal Overhauser field $B_N$
\cite{Salis2009}. The depolarization peak saturates already for relatively small nuclear polarization, but the Hanle
satellites can easily be used as a sensitive probe of $B_N$, in particular for large polarizations and down to the
lowest temperatures.

Employing a novel, all-electrical pump-probe cycle, we measure the nuclear spin relaxation time using the Hanle
satellites over a broad temperature range from $100\,$mK to $20\,$K. At high temperatures $T>1\,$K, the measured $T_1$
times are in good agreement with NMR data in similar systems \cite{Lu2006,Kaur2010}. At the lowest temperatures $T\sim
100\,$mK, we observe $T_{1}$ times as long as 3 hours. Interestingly, despite metallic resistivity of the spin valve
devices, we find that the temperature dependence of the nuclear spin relaxation rate follows a power law $T^{0.6}$
rather than the Korringa law $T_1^{-1}\propto T$ \cite{Korringa1950} naively expected for a metallic system. However,
we argue that the combination of disorder and electron-electron interactions \cite{AAreview} in a degenerate
semiconductor not far from the metal-insulator transition can lead to a deviation from the standard Korringa law within
the hyperfine-mediated nuclear spin relaxation mechanism applicable here. Further, the effects of inhomogeneities in
the nuclear spin ensemble are investigated.

\begin{figure}[tpb]
\includegraphics[width=8.5cm]{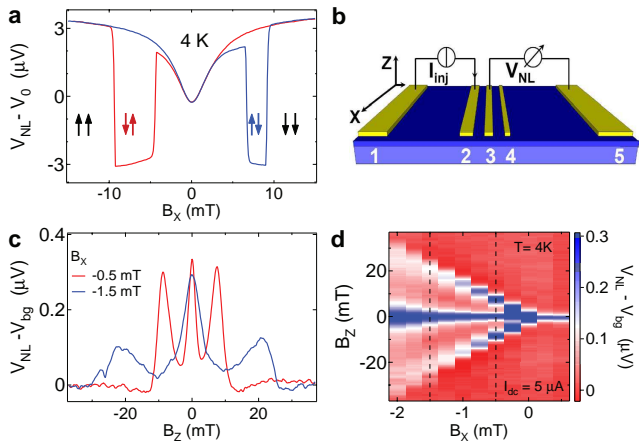}
\caption{\label{fig:1}\textbf{Signatures of nuclear spin polarization in spin valve and Hanle measurements.}
\textbf{a,} Spin valve measurements at 4\, K showing the non-local voltage $V_{NL}$ versus $B_X$ applied along the Fe
bars. An offset $V_0$ is subtracted. Red curve is negative sweep direction, blue positive. Each pair of bold arrows
designates the detector and injector magnetization direction, respectively. \textbf{b,} Device and measurement
schematic. \textbf{c} and \textbf{d,} Hanle measurements at 4\, K. Displayed is $V_{NL}$ (in \textbf{d} as color scale)
versus $B_Z$ applied perpendicular to the sample plane for various $B_X$. The same parabolic background is subtracted
for all $B_Z$ sweeps ($0.34\,$mT/s). Dashed lines correspond to cuts shown in \textbf{c}. We use the satellite peaks
seen for $B_X<0$ as a measure of $B_N$, see text.}
\end{figure}

The spin valve device used in this study consists of five 6\,nm thick Fe bars on a c(4x4) reconstructed surface of a
1\, micron thick epilayer of \textsl{n}-doped GaAs, see Fig.\,\ref{fig:1}b. The three contacts in the center have
widths of 6, 2 and 1$\, \mu$m and edge-to-edge distances of 3 and 4.5$\, \mu$m, respectively. A 15\, nm thick highly
doped surface layer allows efficient spin injection and detection across the Schottky barrier formed at the
metal-semiconductor interface. Within the subsequent 15\, nm, the Si donor density is reduced to the epilayer charge
carrier density of $n=5\,\times\, 10^{16}\, cm^{-3}$. Further device preparation details are described in
\cite{Salis2009}.

The measurement scheme used drives currents from the injector 2 to the 100$\, \mu$m distant contact 1 while a non-local
voltage $V_{NL}$ is measured between contacts 3 and 5 outside the charge current path (see Fig.\,\ref{fig:1}b).
$V_{NL}$ is detected via a lock-in technique using a small ac-modulation on top of a dc-injection current. Electron
spin polarization is efficiently injected into the semiconductor \cite{Zhu2001,Lou2007}, where it diffuses away from
the place of injection. The electron spin diffusion length $\ell_{D}$ is several microns, larger than the distance to
the detector contact \cite{Salis2009}. The measurements are performed in a dilution refrigerator equipped with a
home-built 3-axis vector magnet, allowing us to determine the magnetization direction of the iron bars to better than
$\mathrm{1°}$ by rotating the magnetic field during continued spin valve measurements.

Spin valve measurements are shown in Fig.\,\ref{fig:1}a. Clearly distinguishable voltage levels for parallel and
anti-parallel alignment of injector and detector magnetization are apparent. Further, the non-local spin signal appears
suppressed around $B_X\sim0$ \cite{Lou2007}, as the electron spins start to precess and depolarize due to $B_N\neq0$
\cite{Salis2009}. When $B_N$ is strong enough to completely wash out the polarization of electron spins on their
diffusive path to the detector, the depolarization dip saturates and becomes insensitive for larger $B_N$. Therefore,
the depolarization dip cannot easily serve as a sensitive detector for large nuclear spin polarizations.

To overcome this limitation, we turn to Hanle measurements \cite{Johnson1985, Jedema2002}, where an additional external
field $B_{Z}$ is scanned perpendicular to the sample plane, see Fig.\, \ref{fig:1}c and d. The injected electron spins
precess, diffuse and dephase in the perpendicular field $B_Z$. Around $B_Z \sim 0$, the spins do not precess much and
reach the detector mostly unchanged, giving a peak in the spin signal (Hanle line-shape). For the following, both
injector and detector were initialized along the positive x-direction and are kept parallel throughout. For $B_X<0$,
two satellite peaks appear (Hanle satellites) \cite{Salis2009} as seen in Fig.\, \ref{fig:1}c and d, displaying a
(partial) recovery of the spin signal and hence a suppression of spin precession and dephasing
\cite{Farah1998,Epstein2002}. The satellite peaks thus indicate that $B_{Z}$ is effectively canceled by the z-component
of $B_N$.

Theory estimates the Overhauser field $\textbf{B}_N$ in steady-state and for large $B_{ext}=|\textbf{B}_{ext}|\gg B_L$,
where $B_L\sim1\,$mT is the rms field seen by a nuclear spin, as \cite{Paget1977,Strand2003}:
\begin{equation}
\textbf{B}_{N} = b_{N}^{0}\frac{B_X S_X}{B_{ext}^{2}}\textbf{B}_{ext} , \label{eq:1}
\end{equation}
where $b_{N}^{0}$ is the maximum nuclear field obtainable with DNP and $|S_X|\leq 1$ is the resident electron spin
polarization along the x-direction. Following eq.\,(\ref{eq:1}), $\textbf{B}_N$ is either parallel or antiparallel to
$\textbf{B}_{ext}$ and in particular, $\textbf{B}_{N}$ points opposite $\textbf{B}_{ext}$ for $B_{X}<0$ and $S>0$.
Thus, the external field \emph{vector} $\textbf{B}_{ext}$ can cancel the Overhauser \emph{vector} $\textbf{B}_N$. This
leads to a recovery of the electron spin polarization and appearance of the satellite peaks, which can therefore be
used to measure $\textbf{B}_{N}$ \cite{Meier, Epstein2002, Salis2009}. For $B_Z\gg B_X,B_Y$, we have $B_N \sim B_{Z}$
on the satellite peak. Nuclear fields achieved here are $\sim 50\, $mT, about one percent of the $\sim5\, $T for fully
polarized nuclei in GaAs.

The steady-state situation assumed in eq.\,(\ref{eq:1}) is not typically reached in our measurements, since the ramp
rates employed are fast compared to the time scales of the nuclear spins. %, as we will see.
Performing a time average
$\langle \cdot \rangle_t$ of the magnitude of eq.\,(\ref{eq:1}) gives \cite{Salis2009}
\begin{equation}
\langle B_N \rangle_t \approx b_{N}^{0}B_{X}S_X\left\langle\frac{1}{B_{ext}}\right\rangle_t. \label{eq:2}
\end{equation}
Note that the largest contributions to $\langle \cdot \rangle_t$ arise around $B_Z \sim 0$ and therefore $B_{ext} \sim
B_X$, where the equilibrium $B_N=b_N^0 S_X$ is maximal. The resulting average nuclear field $B_N$ and therefore
satellite peak splitting is linear in $B_X$, as seen in Fig.\,\ref{fig:1}d \cite{Salis2009}. Further, significant
broadening of all three peaks is visible for increasingly negative $B_X$. For the following, we fix $B_{X} = -1.5\, $mT
as a compromise between peak splitting and broadening.

\begin{figure}[tpb]
\includegraphics[width=9cm]{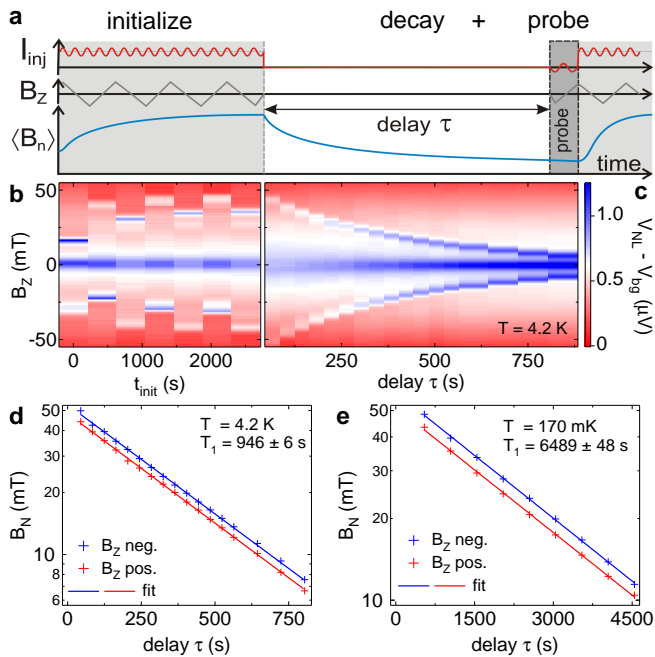}
\caption{\label{fig:2}\textbf{Nuclear spin relaxation determined from Hanle measurements using a pump, decay and probe
cycle.} \textbf{a,} Measurement schematic. \textbf{b,} Initialization: Hanle $B_Z$ sweeps (0.3\,mT/s) as a function of
time with I$_{dc}=$ 20 $\mu$A. Satellite peaks spacing grows and eventually saturates after typically one hour. Trace
begins at $B_Z=+75\,$mT, sweeps to $B_Z=-75\,$mT and then continues to alternate sweep direction. \textbf{c,} Hanle
$B_Z$ probe-traces (0.9\,mT/s) as a function of delay time $\tau$, here starting from a slightly more polarized nuclear
state compared to the end of \textbf{b}. Decay of the satellite peaks is clearly visible. A parabolic background
identical for all delays $\tau$ was subtracted. \textbf{d and e,} Log-plot of Overhauser field $B_{N}$ (crosses) -
extracted from satellite peak positions such as in \textbf{c} - as a function of $\tau$ measured at 4.2\,K in
\textbf{d} and 170\, mK in \textbf{e}. Blue data is from negative-$B_Z$, red from positive-$B_Z$ satellite peaks, with
\emph{single}-exponential fits (solid lines) giving excellent agreement. Very long $T_1$ times are obtained, as
indicated.}
\end{figure}
%\FloatBarrier

Nuclear spin relaxation times $T_1$ are measured using a pump, decay and probe cycle as sketched in Fig.\,\ref{fig:2}a.
First, the nuclear polarization is built up by injecting a large current for efficient DNP while continuously sweeping
$B_{Z}$ back and forth (`initialize'). Nuclear spin polarization is gradually accumulated over time, see
Fig.\,\ref{fig:2}b, eventually reaching a steady state, typically after an hour. As will be shown later, the continuous
sweeping has the effect of homogenizing the nuclear spin system, producing sharper satellites. The asymmetry,
alternating positions and alternating widths of the satellite peaks is a consequence of the time average in
eq.\,(\ref{eq:2}), arising from the alternating sweep directions and efficient DNP when $B_Z\sim 0$ followed by some
decay and homogenization for $|B_Z|\gg0$. After initialization, the nuclear polarization is allowed to decay for a time
$\tau$ with $I_{dc,ac}=0$ and $B_{Z}$ ramped to zero (`decay', keeping $B_X = -1.5\,$mT fixed). Subsequently, a fast
Hanle scan is performed with only a small ac-current (no dc-current) applied to avoid further DNP (`probe'). Repeating
this cycle for various delays $\tau$, data sets reflecting the decay of the nuclear Overhauser field $B_N$ over time
are obtained, as shown in Fig.\,\ref{fig:2}c. Note that the overall measurement time is 10-30 hours due to the very
long nuclear time scales and repeated initialization for each $\tau$.

By fitting Lorentzians to the satellite peaks, we obtain $B_N$ (peak position) as a function of $\tau$, as shown
(crosses) in Fig.\,\ref{fig:2}d at $T=4.2\, $K and Fig.\,\ref{fig:2}e at $T=170\, $mK, for both positive (red) and
negative (blue) $B_Z$ satellites. The small difference between the two satellite positions is again a result of
eq.\,(\ref{eq:2}) and the sweep direction. From exponential fits we get excellent agreement with the data and nearly
identical $T_1$ times for the two satellites. We note that $B_N(\tau)$ clearly follows a single-exponential decay over
the measurable range of $B_N$ (about one order of magnitude in $B_N$) at all temperatures. In addition to the decay of
$B_N$ we observe sharpening of the satellites with increasing delay times, indicating increasing homogeneity of the
nuclear spins with time. At high temperatures $>1\,$K, the $T_1$ times obtained here are in good agreement with
previous $T_1$ measurements by NMR on GaAs with similar doping \cite{Lu2006,Kaur2010}.

\begin{figure}[tpb]
\includegraphics[width=8.5cm]{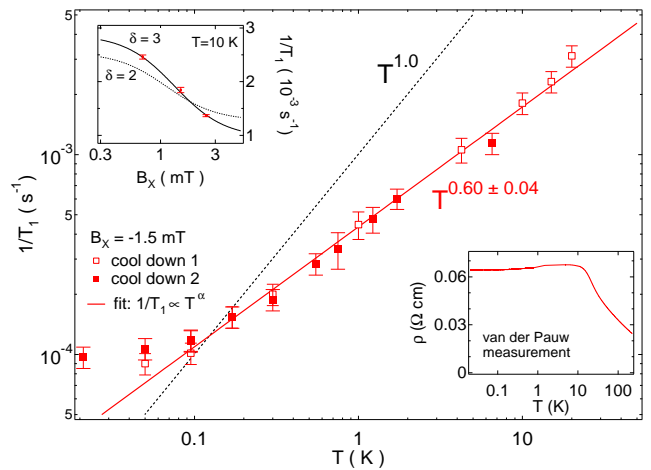}
\caption{\label{fig:3}\textbf{Temperature dependence of the nuclear spin relaxation rate.} $T_1^{-1}(T)$ is shown
between $20\, $mK and $20\,$K measured in two consecutive cool downs (open and solid squares) on the same sample, all
for $B_X = -1.5\,$mT. Error bars are estimated from repeated measurements under identical conditions. The solid (red)
line is a power law fit $1/T_{1} \propto T^{\alpha}$ with $\alpha=0.6\pm0.04$ (excluding the data below $100\,$mK from
the fit). As a comparison, the slope of the Korringa relaxation law $1/T_{1} \propto T$ applicable in metals is added
(dashed line). Upper inset: $B_X$ dependence of the nuclear T$_{1}$-rate at T=10 K with theory fit (dashed curve) to
eq.\,(\ref{eq:3}), see text. Lower inset: temperature dependence of the resistivity from van der Pauw measurements on
the same GaAs wafer, indicating metallic behavior for $T<10\,$K.}
\end{figure}
%\FloatBarrier

The temperature dependence of the nuclear spin relaxation rate is shown in Fig.\,\ref{fig:3} on a log-log plot for two
cool downs (open and closed squares) of the same sample. Measurements of a second sample (not shown) fabricated on the
same wafer give very similar results. Both ac and dc currents were experimentally chosen to avoid self-heating.
However, in the refrigerator we used, sample temperatures saturate around $100\, $mK due to insufficient
thermalization, causing the relaxation rates to saturate at $100\, $mK. Nevertheless, at the lowest temperatures, very
long $T_1$ times exceeding 3 hours are found. Motivated by the approximately linear temperature dependence in the
log-log plot ranging over two orders of magnitude in $T$, we fit a power law $1/T_{1} \propto T^{\alpha}$ for  $100\,
$mK$\,\leq T \leq 10\, $K and find $\alpha=0.6 \pm 0.04$. The data points at $T>10\,$K are excluded from the fit due to
well known phonon contributions above $10\,$K \cite{Lu2006}.

In simple metals, the Korringa law $1/T_1\propto T$ is expected \cite{Korringa1950}. Indeed, this is observed in much
higher doped bulk GaAs (n = 2$\times 10^{18}\,$cm$^{-3}$) \cite{Lu2006} measured with NMR above $1\, $K. These $T_1$
data from Ref. \cite{Lu2006} are added to Fig.\,\ref{fig:3} for comparison (dashed line, extrapolated below $1\,$K).
This clearly contrasts with our results, which show a sub-linear temperature dependence extending well into the
previously unexplored temperature range below $1\,$K.

To shed more light on the unexpected temperature dependence of $T_1^{-1}$, we first investigate the resistivity $\rho$.
The lower inset of Fig.\,\ref{fig:3} shows $\rho(T)$ from van der Pauw measurements done on separate samples from the
same wafer. A weak $T$ dependence with metallic behavior ($d\rho / dT > 0$) is seen for $T<10\,$K, as expected for the
present doping of 5$\, \times\, 10^{16}\,$cm$^{-3}$, above the well-known metal-insulator transition (MIT) in GaAs at
$\sim 2 \times 10^{16}\,$cm$^{-3}$ \cite{Rentzsch1986,Shklovskii1984}. The charge density at $4\,$K is the same as at base temperature
within measurement error. Further, a perpendicular magnetic field at low $T$ has no significant influence on the
resistivity. Therefore, though the density is only a factor 2.5 above the MIT, the resistivity data shows clear
metallic behavior, lacking any hints of incipient localization. In addition, we have confirmed that the highly doped
surface layer does not significantly contribute to transport  apart from facilitating the Fe contacts. The interaction
parameter $r_{S} = E_{C}/E_{F}$ is about $0.6$, with Fermi energy $E_{F} = 7.4\,$meV and average Coulomb energy
$E_{C}=4.1\,$meV, indicating that the samples are approaching the interacting regime $r_S \gtrsim 1$. Further, disorder
is quite strong: $k_F \ell \sim 1.7$, with Fermi wavelength $\lambda_F= 55\,$nm and transport mean free path $\ell =
15\,$nm for $T<10\,$K. Therefore, the epilayer behaves like a degenerately doped semiconductor showing clear metallic
behavior, in the somewhat interacting and strongly disordered regime.

We now discuss the temperature dependence of the nuclear spin relaxation rate. First, we exclude phonon contributions
since these have been shown to be relevant only well above $10\,$K and would result in a quadratic temperature
dependence $T_1^{-1} \propto T^2$ \cite{Lu2006,McNeil1976}. Also, nuclear spin relaxation by paramagnetic impurities is
know to be very weak in doped GaAs \cite{Lu2006}. Next, we consider nuclear spin diffusion out of the $1\,\mathrm{\mu
m}$ thick epilayer. This process is in principle temperature independent in the regime applicable here, but combined
with some other relaxation channel -- e.g. $T_1^{-1}\propto T^{\alpha}$ with $\alpha>0$ -- it could become dominant at
low $T$. However, the nuclear spin diffusion length $\ell_{N} = \sqrt{D_{N}T_{1}}$ with $D_{N}$=
$10^{-13}\,$cm$\mathrm{^{2}}$/s \cite{Paget1982} is smaller than the epilayer thickness, even at low $T$. Furthermore,
this random-walk contribution is inconsistent with the clear single-exponential decay of $B_N(\tau)$ which we find for
all temperatures, also making it unlikely that the observed low-$T$ saturation of $T_1^{-1}$ is caused by nuclear spin
diffusion. Therefore, we estimate that nuclear spin diffusion is negligible here.

Next, we consider the hyperfine contact interaction as a possible nuclear spin relaxation mechanism. In non-degenerate
semiconductors, the Fermi energy is well below the conduction band edge, the mobile charge carriers follow a Boltzmann
distribution, and the nuclear spin relaxation rate is $T_1^{-1}\propto \sqrt{T}$ \cite{Abragam}, not far from the
measured $T_1^{-1}\propto T^{0.6}$. However, since the measured resistivity $\rho(T)$ does not display a thermally
activated behavior expected for a non-degenerate semiconductor, this mechanism is most likely not applicable for a
homogeneous system. In presence of spatial inhomogeneities with non-degenerate (low density) and degenerate (high
density) electron regions, a combination of Korringa ($T_1^{-1}\propto T$) and Boltzmann relaxation ($T_1^{-1}\propto
\sqrt{T}$) could in principle lead to the observed $T_1^{-1}\propto T^{0.6}$. However, for the low-density regions to
remain in the Boltzmann regime down to $\sim 100\,$mK, the density would have to be several orders of magnitude lower
than bulk, making this unlikely.

To learn more about the mechanism of nuclear spin relaxation, we investigate the $B_{X}$ dependence of $T_1^{-1}$,
shown in the upper inset of Fig.\,\ref{fig:3} at $10\, $K. Note that $B_Z=0$ during the decay cycle of the $T_1$
measurement. A clear reduction of relaxation rates is seen for increasing $B_X$, as expected for applied fields
comparable with $B_L$, which is the local rms field acting on each individual nuclear spin, including nuclear
dipole-dipole fields and electronic Knight fields. The theoretically expected rate is \cite{Anderson1959, Hebel1959,
Abragam}
\begin{equation}
T_{1}^{-1}(B) = a \frac{B^2+\delta(5/3) B_{L}^2}{B^2+(5/3) B_{L}^2} \ , \label{eq:3}
\end{equation}
with large-field rate $a=T_{1}^{-1}(B\gg B_L)$. Note that the zero-field rate $T_1^{-1}(B=0)=\delta a$ and the
correlation parameter $\delta$ is ranging from 2 for uncorrelated to 3 for fully spatially correlated fields $B_L$.
Independent measurements give a very small $B$-field offset $<0.1\,$mT, which we assume to be zero here. We perform a
fit constraining $2\leq \delta \leq 3$ and obtain $\delta = 3.0 \pm 0.3$, $B_L = 1 \pm 0.2\,$mT and
$a=(9.6\pm1.5)\times \,10^{-4}\,\mathrm{s^{-1}}$. Further, the dashed theory curve shows a best fit with $\delta = 2$
held fixed, which is clearly inconsistent with the present data. Therefore, $B_L$ is spatially highly correlated
($\delta = 3$) with local field $B_L$ larger than estimated nuclear dipole-dipole fields alone. This suggests an
electronically induced hyperfine Fermi contact interaction mechanism causing nuclear spin relaxation, due to extended
electrons on a length scale much larger than the lattice constant.

For metals or degenerate semiconductors, the hyperfine contact interaction results in a nuclear spin relaxation rate in
the free-electron approximation given by \cite{Abragam} (Korringa mechanism)
\begin{equation}
\frac{1}{T_{1}} = \frac{256\pi^3}{9\hbar} \frac{\gamma _{n}^{2}}{\gamma _{e}^{2}}\ n^2 |\phi(0)|^{4} \chi^{2}\cdot
k_{B}T \ , \label{eq:4}
\end{equation}
with gyromagnetic ratio $\gamma_n$ of the nuclei and $\gamma_e$ of the electrons, electron spin susceptibility $\chi$
and electron density $n\,|\phi(0)|^{2}$ at the nuclear site. This mechanism is consistent with the measured $B_X$
dependence indicating strongly correlated fields $B_L$ and the observed single-exponential decays of $B_N(\tau)$.
Beyond the free-electron model, $T_1^{-1}$ would need to be calculated properly including the combined effects of
disorder and interactions \cite{AAreview, Lee2000, Shastry1994} not far from the MIT. In lack of a $T_1^{-1}$ theory in
this regime, naively, a corresponding renormalized electron spin susceptibility appears in eq.\,(\ref{eq:4}) and can
develop a temperature dependence \cite{Fulde1968, Finkelshtein1984, Castellani1986,Belitz1991}. This was also seen in
other semiconductors for $n\gtrsim n_c$ above but not far from the MIT \cite{Bhatt1986} or the Stoner instability.
Here, $\chi\propto T^{-\beta}$ with $\beta=0.2\pm0.02$ would be required to result in $T_1^{-1}\propto T^{0.6}$ as
measured, assuming no other T dependencies in eq.\,(\ref{eq:4}). Such electronic correlations and $\chi(T)$ are
potentially caused by disordered, interacting mobile electrons, RKKY like effects or spin orbit coupling. In addition,
some fraction of localized electrons in the tails of the impurity band can contribute to nuclear spin relaxation
\cite{Lu2006} and can further cause exchange correlations with delocalized electrons, such as the Kondo effect and
hierarchical formation of spin singlets \cite{Lu2006,Lakner1994,Bhatt1986}. The density dependence of $\beta$ would be
interesting to investigate, indeed, but is beyond the scope of this study. As the free-electron regime is approached
for increasing densities $n\gg n_c$, one would expect $\beta\rightarrow0$, as already reported for larger densities
\cite{Lu2006}.

\begin{figure}[tpb]
\includegraphics[width=8.5cm]{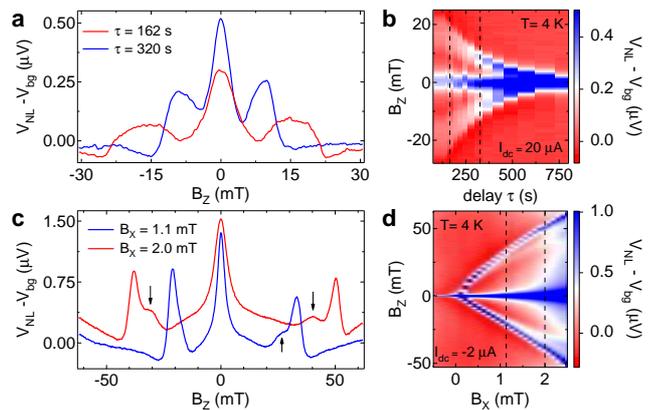}
\caption{\label{fig:4}\textbf{Effects of nuclear spin inhomogeneities.} \textbf{a and b,} are Hanle probe measurements
from a pump, decay and probe cycle as in Fig.\,\ref{fig:2} but here with initialization done at $B_Z=0$ fixed rather
than sweeping. Though the satellite peaks are obviously much broader (than e.g. Fig.\,\ref{fig:2}), indicating stronger
nuclear inhomogeneities, very similar nuclear relaxation times are extracted. \textbf{c and d,} In slow Hanle
measurements ($0.25\,$mT/s), additional satellite peaks become visible (arrows in \textbf{c}) for $B_{X}>0.8\,$mT, see
text.}
\end{figure}
%\FloatBarrier

Finally, we investigate nuclear spin inhomogeneities apparent in Hanle measurements. Fundamentally, the decay of the
electron spin polarization over the space between injection and detection contact can lead to spatial decay and
inhomogeneity in the induced nuclear spin polarization and corresponding satellite peak broadening. However, the
electron spin diffusion length is larger here than the injector-detector distance \cite{Salis2009}. Further, given
sufficient time, nuclear spin relaxation (and potentially diffusion) can counteract polarization imbalances, making the
nuclear inhomogeneities sensitive to the details of the DNP scheme such as sweep times and $B_Z$ field conditions.

Since the strongest DNP occurs around $B_Z\sim0$ (see eq.\,(\ref{eq:2})), we fix $B_Z=0$ during initialization (rather
than sweeping it), followed by decay and probe cycles as for the $T_1^{-1}$ measurements previously. As seen in
Fig.\,\ref{fig:4}a and b, this results in significantly broadened satellite peaks compared to Fig.\,\ref{eq:2}c,
consistent with the absence of a nuclear spin homogenization mechanism as discussed before in Fig.\,\ref{fig:2}b.
Despite the additional broadening, however, the extracted nuclear spin relaxation times (not shown) are identical
within the experimental error to the sweeping $B_Z$ initialization in Fig.\,\ref{fig:2}. Further, clear sharpening of
satellite peaks is seen in Fig.\,\ref{eq:4} with progressing delay $\tau$, consistent with our previous observations in
Fig.\,\ref{fig:2}b and c.

Additional satellite peaks are present in Hanle measurements, see Fig.\,\ref{fig:4}c, clearly visible but smaller and
more broadened than the main satellites. Note that double satellites appear whenever the Hanle satellites are well
enough separated and sharp enough, irrespective of current direction and sign of $B_X$, here shown for $B_X>0$ and
reversed current in Fig.\,\ref{fig:4}d. The additional satellites have also been observed on a second device and seem
to suggest two distinct species of electron and/or nuclear polarization regions between injector and detector. Drifting
electrons will also feel the spin orbit interaction, which might alter the effective magnetic field they are exposed
to. Further studies are needed to elucidate these additional peaks.

In summary, the presented measurements on GaAs spin valves show the importance of the nuclear spin environment, which
can easily be manipulated all-electrically via DNP at temperatures below 20 K. Extremely long nuclear spin relaxation
times exceeding 3 hours are seen. The nuclear spin relaxation rate has a rather weak, power-law temperature dependence
$T_1^{-1}\propto T^{0.6}$, deviating from the standard Korringa law $T_1^{-1}\propto T$ in metals, but not inconsistent
with nuclear spin relaxation via the hyperfine contact interaction in the interacting, disordered electron regime
currently lacking theory. Inhomogeneities in the nuclear spin polarization were investigated, and appeared not to
affect the spin relaxation rate.

%\bibliography{NatPhys_ManRef_DMZ}

\textbf{Acknowledgments}
\newline This work was supported by the University of Basel, the Swiss Nanoscience Institute (SNI), Swiss NSF, NCCR nano
and NCCR QSIT. We are very thankful for discussions with Bernd Braunecker, Daniel Loss and Mitya Maslov.
\newline
\textbf{Author Contributions}
\newline A.F., G.S. and S.F.A. designed and fabricated the devices. D.K. and D.M.Z. performed the experiments. All
authors contributed to discussions during execution and interpretation of the experiment. D.K. and D.M.Z. wrote the
manuscript with feedback from A.F., G.S. and S.F.A.
\newline
\textbf{}
\newline The authors declare that they have no competing financial interests.

\end{document}